\documentclass[sigconf]{acmart}

\AtBeginDocument{%
  \providecommand\BibTeX{{%
    \normalfont B\kern-0.5em{\scshape i\kern-0.25em b}\kern-0.8em\TeX}}}

\copyrightyear{2023}
\acmYear{2023}
\setcopyright{rightsretained}
\acmConference[CHI '23]{Proceedings of the 2023 CHI Conference on Human Factors in Computing Systems}{April 23--28, 2023}{Hamburg, Germany}
\acmBooktitle{Proceedings of the 2023 CHI Conference on Human Factors in Computing Systems (CHI '23), April 23--28, 2023, Hamburg, Germany}\acmDOI{10.1145/3544548.3580651}
\acmISBN{978-1-4503-9421-5/23/04}
\acmDOI{10.1145/3544548.3580651}

\usepackage[utf8]{inputenc}
\usepackage[T1]{fontenc}

\usepackage{xcolor}
\usepackage{xspace}
\usepackage{color}
\usepackage{multirow}

\definecolor{applegreen}{rgb}{0.55, 0.71, 0.0}
\definecolor{amber}{rgb}{1.0, 0.49, 0.0}
\definecolor{fuschia}{rgb}{0.57, 0.36, 0.51}
\definecolor{ashgrey}{rgb}{0.7, 0.75, 0.71}
\definecolor{beaver}{rgb}{0.62, 0.51, 0.44}
\definecolor{teal}{rgb}{0.0, 0.5, 0.69}
\definecolor{blue-violet}{rgb}{0.54, 0.17, 0.89}

\newcommand{\nextitem}{\par\hspace*{-1mm}\textbullet\hspace*{\labelsep}}

\begin{document}


\title[How does HCI Understand Human Agency and Autonomy?]{How does HCI Understand Human Agency and Autonomy?}

\author{Daniel Bennett}
\email{pavementsands@gmail.com}
\orcid{0000-0002-9330-5529}
\affiliation{\institution{Aalto University} \city{Espoo} \country{Helsinki}}
\author{Oussama Metatla}
\orcid{0000-0002-3882-3243}
\affiliation{\institution{University of Bristol} \city{Bristol} \country{United Kingdom}}
\author{Anne Roudaut}
\orcid{0000-0003-3082-4564}
\affiliation{\institution{University of Bristol} \city{Bristol} \country{United Kingdom}}
\author{Elisa D. Mekler}
\orcid{0000-0003-0076-6703}
\affiliation{\institution{IT University of Copenhagen} \city{Copenhagen} \country{Denmark}}

\renewcommand{\shortauthors}{Bennett, Metatla, Roudaut \& Mekler}


\begin{abstract}

Human agency and autonomy have always been fundamental concepts in HCI. New developments, including ubiquitous AI and the growing integration of technologies into our lives, make these issues ever pressing, as technologies increase their ability to influence our behaviours and values. However, in HCI understandings of autonomy and agency remain ambiguous. Both concepts are used to describe a wide range of phenomena pertaining to sense-of-control, material independence, and identity. It is unclear to what degree these understandings are compatible, and how they support the development of research programs and practical interventions. We address this by reviewing 30 years of HCI research on autonomy and agency to identify current understandings, open issues, and future directions. 
From this analysis, we identify ethical issues, and outline key themes to guide future work. We also articulate avenues for advancing clarity and specificity around these concepts, and for coordinating integrative work across different HCI communities.
\end{abstract}




\maketitle

\section{Introduction}

Autonomy and agency have been fundamental concepts in Human-Computer Interaction (HCI) since its inception. They have been seen as guiding values to inform design \cite{friedman_value-sensitive_1996,friedman_user_1996}, as sources of pleasurable and meaningful user experiences \cite{hassenzahl_designing_2013,peters_designing_2018}, and as basic factors in users’ sense of ownership over outcomes \cite{C18}. Developments in recent years have only made these concepts more pertinent. Issues including the growth of persuasive technologies \cite{hamari_persuasive_2014}, interaction with intelligent systems \cite{Calvo2020}, and the tight integration of technologies into our lives and bodies \cite{cornelio2022sense,mueller_next_2020} all raise implications for human autonomy and agency.
As such, agency and autonomy stand out as potential \textit{boundary objects} for 
HCI. 
Boundary objects are constructs which hold important roles across multiple communities, and which function to
coordinate the perspectives of these communities \cite{leigh_star_this_2010}.
Constructs which function in this way have been seen as key in developing shared problem definitions, and generating robust knowledge across multiple domains \cite{nuchter_concept_2021}. They are characterised by a degree of 
flexibility
in meaning, which allows them to serve the communicative and informational needs of each community involved \cite{nuchter_concept_2021}, without relying on strict consensus \cite{leigh_star_this_2010}. 
At the same time, they must balance this flexibility with robustness in meaning, which allows them to maintain their identities across communities, and thereby support negotiation of understanding between these communities \cite{velt_translations_2020, nuchter_concept_2021}.
Turning back to the concepts of agency and autonomy in HCI: their ubiquity and longstanding importance across a range of communities suggests that they are flexible enough to fulfil this function.
However, it is currently unclear 
to what extent they balance this flexibility with robustness in meaning and identity. 
Agency and autonomy are notoriously complex concepts: 
They are sometimes treated as synonymous, and sometimes held to have distinct, albeit deeply entangled, meanings \cite{cummins_agency_2014}. As such it is not clear whether 
sufficient overlap in understandings are currently found across different approaches in HCI. Nor is it clear whether there might be opportunities to clarify mutual understandings so that communities may work together and learn from one another.

Recent research has aimed to clarify understandings 
of these concepts in particular domains: in the autonomy experience of Multiple Sclerosis sufferers \cite{C70}; 
the immediate sense-of-agency over discrete interaction gestures \cite{limerick_experience_2014}; 
and in human-AI interaction \cite{burr_supporting_2020,Calvo2020}. However, 
it is not clear how these accounts might relate to one another, 
nor where each sits within the broader landscape of research on agency and autonomy in HCI.
At present it is unclear how autonomy and agency figure in HCI research, how researchers across different HCI communities currently understand autonomy and agency, and the degree to which these understandings are compatible.

To address this, we present findings from a systematic literature review, encompassing 161 papers over 32 years of research (1990 to 2022),
taking stock of notions of autonomy and agency in HCI literature. Our contribution is three-fold: First, we show that autonomy and agency currently figure not as effective boundary objects in the HCI landscape, but as vague \emph{umbrella constructs} 
\cite{tractinsky_usability_2018} -- broad concepts, subsuming a surprising diversity of understandings and theoretical underpinnings. We find that ``autonomy'' and ``agency'' are often used interchangeably, and given a wide range of different meanings. 
These include
the \textit{implicit} \textit{experience} that our actions are responsible for outcomes \cite{C12}; the \textit{material} influence we have on a situation \cite{C78}, and the \textit{experience} that outcomes are congruent with one's values \cite{C141}. 
Despite this breadth of understandings,
we find that few works give specific definitions of autonomy and agency, nor examine the relationships of these concepts to expected outcomes. 

Second, we outline 
ways in which HCI could
move beyond this umbrella approach and develop agency and autonomy as well-functioning boundary objects. 
Specifically, we identify four aspects which characterise approaches to agency and autonomy in HCI. 
Works were distinguished by their focus on 1. issues of \textit{self-causality} and personal \textit{identity}; 2. the \textit{experience} and the \textit{material} expression of autonomy and agency; 3. particular \textit{time-scales}; and 4. emphasis on \textit{independence} and \textit{interdependence}.
These aspects can help future HCI research identify relevant prior work, draw meaningful distinctions within understandings of autonomy and agency, and clarify approaches to their support.  

Finally,
we use these four aspects of autonomy and agency to articulate open questions and future directions for work in HCI. We outline avenues to clarify issues of autonomy and agency and their relationships to valued outcomes, and capitalise on commonalities between research in different areas. We also articulate ethical challenges around technology and human agency and autonomy. 

\section{Background}


In technology research ``autonomy'' is often linked to robotics and AI \cite[e.g., `autonomous vehicles',][]{meschtscherjakov_interacting_2018}, and ``agent'' can refer to software agents which carry out tasks, 
and coordinate their actions with people \cite{nwana_software_1996}. 
Our review focuses exclusively on \emph{human} autonomy and agency. Work on autonomous and agentic technologies, and 
agency in non-humans, will only be discussed insofar as it bears on human agency and autonomy. 

Autonomy and agency are closely related terms, though they are distinct in etymology, and often --- but not always --- in usage.
``Autonomy'' derives from the Greek \emph{autos}, meaning ``self'', and \emph{nomos}, meaning ``custom or law'' \cite{oshana_personal_2016},  together taken to mean ``self-governance'' \cite{killmister_taking_2018}. By contrast, ``agency'' has its roots in the Latin \emph{agere}: ``to set in motion, drive forward; to do, perform'' \cite{douglas_harper_origin_2022}, reflecting an emphasis on self-causality. 
While modern scientific and philosophical usage of the terms often follows these etymologies \cite[e.g.,][]{buss_personal_2018,killmister_taking_2018, luck_formal_1995,abrams_autonomy_1998,vallerand_intrinsic_2002}, the distinction between performing and governing is not always easily drawn in everyday life. Recent philosophy has emphasised that while distinct, the terms are tightly entangled \cite{cummins_agency_2014}, and elsewhere the terms are often treated as (near) synonymous: In HCI, for example, Peters et al. recently defined autonomy as ``feeling agency, acting in accordance with one’s goals and values'' \cite[][p.~2]{peters_designing_2018}. 

This entanglement is also visible in the long history of the terms in philosophy and other disciplines. When Aristotle articulated his account of the `eudaimonic' good life, he gave a central role to \emph{autarkeia}, or self-sufficiency in action and thought \cite{kenny_aristotle_1992}. This emphasis on individualism developed over several centuries, most notably in Kant's culturally definitive account of individual autonomy as the basis of rational agency \cite{kant_cambridge_1996}. Recent philosophy has added complexity to such views: both by emphasising the presence of different orders of agency and autonomy at different timescales \cite{frankfurt_freedom_1988, juarrero_dynamics_1999, barad_meeting_2007}, and by treating individual expressions of agency and autonomy as inherently tied to their socio-material contexts 
\cite{oshana_personal_2016,killmister_taking_2018, barad_meeting_2007}. 
This tension between individual and context is also found in modern positive psychology frameworks such as Self-Determination Theory, which outlines an ``organismic integrative'' approach \cite{ryan_brick_2019}: a complex relationship of mutual constitution between the individual and their social context, in which the individual not only self-governs and acts, but also autonomously integrates aspects of their context into their identity and behaviour \cite{ryan_autonomy_2004}. In such accounts, understanding autonomy or agency is a complex and multi-dimensional matter that requires understanding both causal influence, and how contexts and outcomes inform an individual's identity, \cite{killmister_taking_2018}  goals, and values \cite{ryan_autonomy_2004, killmister_taking_2018}. 
Crucially, this emphasis on the complex entanglement of causation, value, and identity is not only relevant to ``higher order'' motivational and decisional issues. Recent experimental work on low-level sense-of-agency, focusing on the moment-to-moment experience of self-causation, also finds that objective correlates of agency can be impacted by personality factors \cite{hascalovitz_personality_2015}, and by social factors \cite{malik_social_2019}. 

\begin{figure*}
	\centering
	\includegraphics[width=0.6\textwidth]{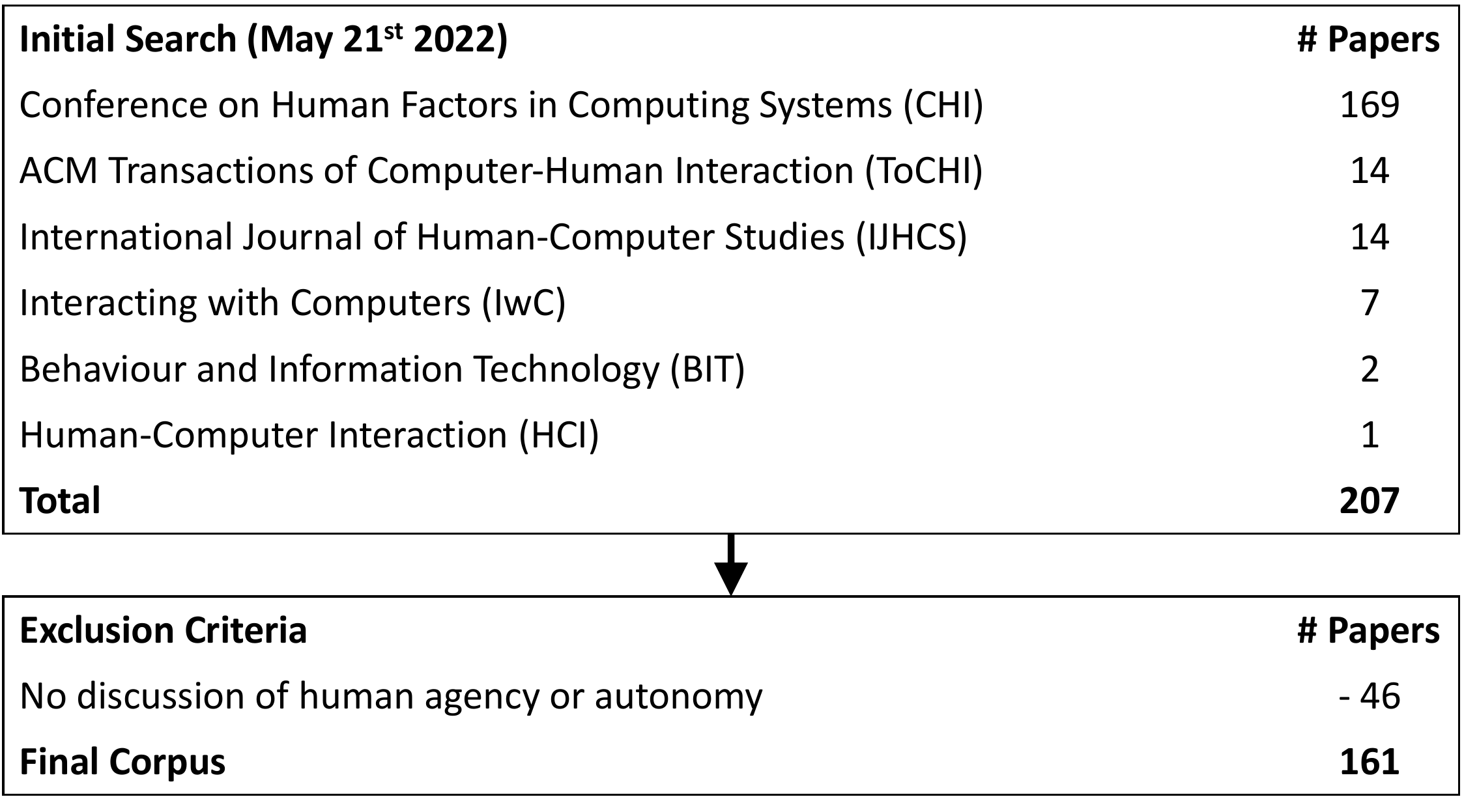}
	\caption{Summary of the literature review procedure.}~\label{review_procedure}
	\vspace{-2em}
\end{figure*}

\subsection{Agency and Autonomy in HCI}

Agency and autonomy have long been focal to HCI scholarship. 
Schneiderman's 1985 \textit{Eight Golden Rules of Interface Design} urged designers to ``support internal locus of control'' \cite[][p.~75]{shneiderman_designing_2010}. 
CHI workshops in 1996 \cite{friedman_user_1996} and 2014 \cite{calvo_autonomy_2014} addressed issues of autonomy in current technology use, covering the risks of undermining human autonomy \cite{friedman_user_1996}, and theory and design strategies related to user autonomy \cite{calvo_autonomy_2014}.
The first of these led to the influential Value-Sensitive Design approach \cite{friedman_value-sensitive_1996}. Organisers of the latter developed the METUX (Motivation, Engagement, and Thriving in User Experience) framework \cite{peters_designing_2018}, which provides guidelines for design for human well-being, and draws attention to the multi-level nature of autonomy.  
Autonomy and agency have figured in a range of work aiming to characterise good user experience (UX) \cite[e.g.,][]{C140,hassenzahl_designing_2013,tuch2013analyzing}
and they remain central concepts in recent strands of HCI research: For example, in Human-Computer Integration, agency is suggested to be central to categorising and understanding situations where ``user and technology together form a closely coupled system'' \cite[][p.~1]{mueller_next_2020,cornelio2022sense}. 

Beyond theoretical work, much recent empirical and design work addresses the question of how and when autonomy and agency should be supported. This work, discussed in greater depth in our literature review below, varies considerably in the kinds of agency and autonomy it addresses -- covering for example, the experience of causing discrete events during minimal interactions \cite[e.g.,][]{C62}; the experience of autonomy in episodes of gaming \cite[e.g.,][]{C141}; 
but also the lived experience and personhood of dementia sufferers \cite[e.g.,][]{C121}; and the material fact of control or influence in tasks \cite[e.g.,][]{C110}. At present it is unclear how such different approaches to autonomy and agency, at different scales of behaviour, relate to one another. Moreover the use of near-identical language across this diversity of cases can make it difficult for HCI researchers to identify work relevant to their particular concerns. 
At the same time, we argue that there is value in 
understanding the interrelation of these diverse approaches: The tight integration of technologies into our bodies, behaviours and lives \cite{mueller_next_2020} has implications for autonomy and agency across multiple levels.
Recent work in philosophy of cognition, for example, has indicated risks for integrated technology whereby delegation to technology at sensorimotor timescales (milliseconds) might be expected to impact on autonomy and agency in decision-making and everyday life \cite{meacham_over-extended_2018, wheeler_reappearing_2018}. To understand such scenarios it is imperative to grasp how different aspects of agency and autonomy relate to one-another.

While recent years have seen repeated calls to deal with autonomy in a nuanced and multi-faceted way \cite[e.g.,][]{calvo_autonomy_2014, C70, hornung_autonomy_2016},
it remains unclear what a multilevel understanding of agency and autonomy, adequate to the range of work in HCI, might look like. 
Some calls for nuance have focused only on
particular domains, such as multiple sclerosis support \cite{C70}. The METUX framework \cite{peters_designing_2018, burr_supporting_2020} offers a less domain specific approach and outlines several ``spheres'' of life into which interactions and their implications can be categorised --- from interaction at the interface, to the user's wider life and society.  
Specifically, the METUX calls attention to the presence of multiple levels of need satisfaction, and provides a heuristic for ``organizing thinking and evaluation'' \cite[p.~7]{peters_designing_2018} in design work. 
However, METUX does not focus specifically on autonomy, and the authors do not relate their account to previous work on autonomy and agency in HCI, nor address the full range of issues presented here. 
Also, the basis for choosing its approach to distinguishing aspects of autonomy issue --- ``spheres'' of life --- is not made clear.
As the results of our literature review will show, a number of other aspects
seem operative in approaches to autonomy and agency, and offer clarity in distinguishing understandings and issues. 
To our knowledge, no work has attempted to survey the breadth of approaches to autonomy and agency in HCI, to understand how they relate, and how they might coordinate and build upon one another. This is the goal of our paper.

\section{Review Method}

This paper aims to give an overview of how HCI researchers have addressed and understood issues of agency and autonomy. To this end, we review a corpus of papers that spans 32 years of HCI research.

\subsection{Source Selection}
We searched the ACM Digital Library on May 21st, 2022, for relevant publications from the CHI conference, ACM Transactions of Computer-Human Interaction (ToCHI), and International Journal of Human-Computer Studies (IJHCS), Behaviour and Information Technology (BIT), and Human-Computer Interaction, as these have previously been considered high-impact venues in HCI \cite{hornbaek_technology_2017}. We searched these venues for the keywords "agency" and "autonomy" in keywords and abstracts. 
This resulted in 207 full papers (see \autoref{review_procedure} for breakdown per publication venue). 

\subsection{Selection of papers for inclusion in the review}
We first reviewed abstracts and titles for all papers for whether they concerned human autonomy or agency. Where abstracts focused on non-human agency and autonomy (e.g., of robots, software agents), we reviewed the full text, excluding it if there was no discussion of human agency and autonomy. Example exclusions include one paper which dealt with the agency of parrots in animal-computer interaction \cite{kleinberger_interspecies_2020}, and another focusing on the calibration of autonomous braking systems in cars \cite{fu_is_2020}. In total, 46 papers were removed in this way, leaving 161 papers for analysis.

\subsection{Coding Procedure}
\label{Coding}

\subsubsection{Developing the Coding Rubric}
We analysed our corpus using a rubric, which we developed through 
thematic analysis \cite{ibrahim_thematic_2012}
of a subset of 77 papers.  This subset was obtained through an initial search on the ACM DL, restricted to 10 years of research, at CHI and ToCHI. The analysis followed the process outlined by Braun and Clarke \cite{braun_using_2006}. Our goal for the rubric was to identify aspects and understandings of agency and autonomy in prior HCI work, whether they operated explicitly or implicitly. We therefore opted for an inductive coding process, which was additionally informed by our reading of the work on agency and autonomy 
discussed in our background section. 
The first author read and coded the full texts of all 77 papers in NVivo, 
focusing on passages dealing with autonomy and agency. 
Codes were generated at sentence and paragraph level, aiming at granular description of how autonomy and agency were addressed.
Through this initial step, 466 distinct codes were identified. These were then thematically grouped into higher-level codes.
Finally, the higher-level codes were collated into 7 categories for the analysis of the wider corpus (see \autoref{tbl_rubric} for descriptions and definitions of terms). At each stage of the thematic analysis, the first author checked that the grouped codes remained descriptive of the accounts in the papers. Also, following recommendations in \cite{ibrahim_thematic_2012}, peer validation was conducted throughout this process through regular meetings among the co-authors to review and clarify coding and grouping decisions.


\subsubsection{Coding of the Full Corpus}
The first author read and analysed the full texts of all 161 papers, using the coding rubric. 
For each of the 7 categories in the rubric, the approaches to autonomy and agency in each paper were categorised, and informative quotes supporting this categorisation were recorded. To calibrate analysis, the first and fourth author independently coded the same, randomly selected subset of 13 papers. Differences were resolved through discussion, with only minor revisions to the coding rubric (i.e., clarifying the definitions of the different time-scales). 
The coding spreadsheets are included as supplementary material.

\begin{table*}[ht]
\caption{Categories of analysis and main findings.  For all "aspects" papers could be coded in multiple categories. See supplementary material for detailed coding for each paper in the corpus.}
\centering
\begin{small}
\setlength{\tabcolsep}{1em}
{
\begin{tabular}{| p{0.75 em} | p{6.75em}| p{13em} | p{25em}| }
\hline
\multicolumn{2}{|p{7.5em}|}{\textbf{Category}}  & \textbf{Description}& \textbf{Details / Definitions}  \\
\hline
\multicolumn{2}{|p{7.5em}|}{Subject Matter}  & What was the domain or subject focus of the paper?& Assigned based on keywords and paper topic. (See additional materials)\\
\hline
\multicolumn{2}{|p{7.5em}|}{Value and Benefits} & What values and benefits did authors associate with autonomy and agency?  & 
    \nextitem \textit{Not clear} (n=57).
    \nextitem \textit{Intrinsically good} (n=21).
    \nextitem \textit{Benefits for the User} (n=90).
    \nextitem \textit{Benefits for Other Stakeholders} (n=23).\\
\hline
\multicolumn{2}{|p{7.5em}|}{Articulating \mbox{Understanding} } 
& How did authors articulate their understanding of agency and autonomy? & 
\nextitem \textit{Explicit definition } (n=58).
\nextitem \textit{Meaning indicated via associative sentences} (n=47).
\nextitem \textit{Other } (n=55). \\
\hline
& \mbox{Self-Causality} \mbox{and Identity}  & Were autonomy and agency described in terms of the user's causal involvement (acting and making decisions), and/or in terms of the user's self and values? & 
    \nextitem \textit{Executing}: personal and causal involvement in tasks, processes, outcomes (n=122).
    \nextitem \textit{Decision}: exercising personal choice, and making decisions about tasks, processes, outcomes (n=99).
    \nextitem \textit{Self-Congruence}: Outcome in line with user's values and goals, regardless of causal involvement (n=41).
    \nextitem \textit{Self-Construction}: Impact on identity, values, goals (n=24).\\
\cline{2-4}
\multirow{4}{*}{\rotatebox{90}{\parbox{10em}{\textbf{Aspects of Autonomy}}{\raggedright}}}  & \mbox{Experience and} \mbox{Materiality} & Did the account focus on the user's experience of autonomy and agency, or on the material enaction of autonomy and agency?& 
\nextitem\textit{Material:} E.g. discussion focusing on the user's \textit{material} ability to effect some outcome, manage self-care, etc. (n=113). 
\nextitem\textit{Experiential:} e.g. discussion focusing on the user's first person experience, sense, or feelings of agency and autonomy 
(n=79).\\
\cline{2-4}
& \mbox{Time-scales} & What were the time-scales of the relevant experiences and activities?  & 
\nextitem \textit{Micro-interaction:} A few seconds or less (n=13)
\nextitem \textit{Episode:} Up to a few hours (n=113)
\nextitem \textit{Life:} Days or above (n=77)
\mbox{We used direct statements of time} where available (n=29), inferred from context where possible, 
and did not code where unclear.\\
\cline{2-4}
& \mbox{Independence or} \mbox{Interdependence} & Did the discussion of agency and autonomy emphasise independence, or interdependence?  & 
\nextitem \textit{Independence:} emphasis on acting, thinking, being independent of others  (n=50).
\nextitem \textit{Interdependence:} emphasis on reliance on others, acting as part of a group, or the social grounds of agency and autonomy 
\mbox{Both of these could coincide} in the same paper (n=54). \\
\hline
\end{tabular}
}
\end{small}
\label{tbl_rubric}
\end{table*}

\section{Results}
In the following, we report our analysis of the 161 papers reviewed. We first summarise the range of subject matters addressed and the benefits and values ascribed to agency and autonomy. We then address how authors articulated their understandings of autonomy and agency, before finally presenting the aspects we found in authors' understandings of agency and autonomy. \autoref{tbl_rubric} provides a summary and guide to the analysis categories.

\subsection{Subject matter}


The papers in our corpus covered a wide range of subject matters, from the design and evaluation of input and control methods \cite[][]{C110}, to issues of everyday living \cite[][]{C135}, family and parenthood \cite[][]{C161}. The largest group dealt with issues of ageing and accessibility (n=33): Here autonomy and agency often concerned how individuals could be supported in retaining personhood \cite[][]{C1}, maintaining social and family ties \cite[][]{C8, C152}, and living independent lives \cite[][]{C135}, a focus also seen in work on parenting (n=6), children (n=6), and in work concerned with HCI4D (n=6). Across all these domains, issues of autonomy and agency commonly focused on material matters, addressing constraints imposed by health, and by social and material conditions. 
The next largest group focused on gaming, social media, and entertainment (n=32). Here, issues of autonomy and agency usually centred on people's experiences: how free, independent, self-governing, and in control they felt during the interaction. 
The remaining papers focused on a range of other subjects; from work (n=17), to input methods (n=6). 

\subsection{Value and Outcomes}




Across our corpus it was evident that autonomy and agency are widely considered desirable qualities in HCI. However, in many papers this value was left implicit or 
unclear (n=57). For example, one paper reported that ``users desire agency and control in all aspects of the system'' \cite[][p.~1270]{C143} but did not elaborate on why this might be. Other papers sought to understand how technologies could support agency and autonomy, but did not address their value \cite[e.g.][]{C65, C71}. In the following, we examine in more detail the reasons \emph{why} agency and autonomy are valued, as well as what beneficial \emph{outcomes} HCI researchers associate them with.

\subsubsection{Agency and Autonomy as Intrinsic Goods}

Several papers (n = 21) indicated that autonomy and agency have \textit{intrinsic} value, in and of themselves. Often this was indicated by explicit reference to autonomy or agency as 
``basic psychological need'' \cite[e.g.,][]{C27}, ``fundamental human need'' \cite[e.g.,][]{C155}, or ``universal need" \cite[e.g.,][]{C38,C140}. 
Lukoff et al., for instance, argue that ``sense of agency matters in its own right. Feeling in control of one’s actions is integral to autonomy, one of the three basic human needs outlined in self-determination theory'' \cite[][p.~3]{C27}. Elsewhere, autonomy was considered a ``right'' drawing on the ``UN Convention on the rights of people with disabilities'' \cite[][p.~3249]{C135}, a ``central ethical value'' \cite[][p.~3]{C32} drawing on Value Sensitive Design, or fundamental to human dignity \cite{C65}.
Other papers hinted at the intrinsic value of agency and autonomy by problematising their absence. Irani and Six Silberman \cite{C2}, for example, explored accounts on Turkopticon that describe crowdworkers as ``exploited cogs in other people’s plans, toiling in digital sweatshops with little creativity or agency'' \cite[][p.~4574]{C2}.

\subsubsection{Beneficial Outcomes of Agency and Autonomy for the User}
The majority (n=90) of reviewed papers located the value of agency and autonomy in their benefits for users. The nature of these benefits, however, varied considerably from work to work: 
Many papers 
valued agency and autonomy as 
key constituents of users' mental and physical well-being and sense-of-self. 
Specifically, agency and autonomy were associated with a slew of positive outcomes for the user, including improved life satisfaction \cite[e.g.,][]{C134}; overcoming barriers to acting or choosing (via e.g. ageing, disability, resources, or other material factors) \cite[e.g.,][]{C19, C45, C121}; and acting as a buffer against adverse effects, such as stress, impaired sleep and diminished social interactions \cite[e.g.,][]{C27}. 
Likewise, notions of agency and autonomy were considered crucial to physiological and physical health \cite[e.g.][]{C25, C34}, supporting better health outcomes \cite{C131}, and contributing to positive developmental outcomes \cite{C15, C46, C57, C114}. 


Next, several papers linked autonomy and agency to good user experience (UX) -- often in a rather general manner \cite[e.g.][]{C80,C38,C18}, with no concrete ties to specific aspects or qualities of experience. Some works were slightly more precise, tying agency and autonomy to technology satisfaction \cite[e.g.][]{C27}, sense of control \cite{C128, C160, C62}, or some measure of game enjoyment \cite{C79,C113,C126}. 
Other works linked greater agency and autonomy to greater sense of meaning for users, 
for instance, by allowing users to create their own meaning in exhibitions \cite{C147} 
and educational settings \cite{C149}. Besides individual meaning, autonomy and agency in interpersonal interactions was associated with more meaningful communication \cite{C13, C25, C59} and 
facilitating group meaning \cite{C11}.


Agency and autonomy were also valued for a range of desirable pragmatic outcomes for the user, from enhanced learning \cite{C125}, to improved privacy and security \cite[e.g.,][]{C41,C72,C150}. 
One paper noted that ``autonomy is like the ``muscle'' of privacy'' \cite[][p.~358]{C123}, and in line with this a number of works 
explored the role of autonomy and agency in supporting privacy \cite{C44,C72,C123,C150}.
Concurrently, some works indicated tensions between agency and autonomy versus safety, security and privacy, e.g., in the case of children or people with diminished capacities. These works \cite{C30, C24, C29, C46} emphasised that granting full agency and autonomy might leave people to pose risks to their own physical or digital safety, and suggest a need to balance these values. 
Finally, 20 papers suggested that user autonomy and agency contribute to desirable practical outcomes in tasks and activities \cite[e.g.,][]{C33,C56,C139,C148},  
such as increased feelings of responsibility \cite{C47}, or 
trust within organisations \cite{C136}. 

\subsubsection{Beneficial Outcomes of Agency and Autonomy for Other Stakeholders}
Beyond the individual user, benefits of autonomy or agency 
could also accrue to social groups, organisations and companies (n=23). In some, but not all cases, these beneficial outcomes appear likely to align with the users' own values and choices --- for example, where autonomy and agency were seen as supporting meaningful sharing in communities \cite{C11, C22} and families \cite{C13, C25, C114, C129}. 
In other instances, organisations and companies benefited from users' autonomy and agency, with mixed outcomes for the user. 
In some works, for instance, organisations benefited from efficiency and motivational outcomes, while individuals also benefited materially in the form of improved health outcomes \cite{C58} financial benefits \cite{C73}, and lifestyle flexibility \cite{C50}. In other cases, the benefits to the user were primarily experiential, while the organisation saw material benefits such as improved work-outcomes, and stronger attachment in games, products and services \cite[e.g.,][]{C90, C50,C94, C139}. 
Finally, in some instances, 
it was not clear that the user gained benefits that they would have chosen for themselves. For example, where organisations granted a ``voice'' to users, while exploiting the appearance of inclusion to provide legitimacy for existing goals \cite{C35, C96}. In another case, workers were afforded autonomy in task scheduling and decision making, but any benefits of this to the worker were counterbalanced by the burden of additional stress, responsibility, unaccounted labour, and skipped toilet breaks \cite{C90}.

\subsection{How Did Authors Articulate Understandings of Agency and Autonomy?}

Only one third of papers in our corpus (n=58) gave explicit definitions of autonomy or agency. Lukoff et al., for example, defined sense-of-agency as ``an individual’s experience of being the initiator of their actions in the world [...] broken down into feelings of agency, that is, the in-the-moment perception of control, and judgments of agency, that is, the post hoc, explicit attribution of an action to the self or other'' \cite[][p.~3]{C27}.
However, even some explicit definitions were rather unspecific (``Human autonomy is the ability `to be one's own person' and impacts well-being'' \cite[][p.~10]{C157}) or tautological (``autonomy ("a sense of freedom and autonomy to play the game as desired'' \cite[][p.~6]{C117}).


In 47 papers, definitions were implied via associative statements. 
For example, the work by da Rocha et al.~\cite{C14} contained a number of statements which together suggested that they used ``autonomy'' to refer to a player's ability to act for themselves without support in gaming sessions (e.g., "The goal is to improve a player’s autonomy and game enjoyment by removing barriers that can hinder their ability to learn a game or engage in gameplay." \cite[][p.~10]{C14}.)
In another 55 papers, the meaning of autonomy or agency was less clear, or rather inconsistent. For instance, Gonçalves et al. described agency in terms of decision making, distinguished from mere execution: ``decision making on the part of both players [...] Avoid the perception that one player is just an “executor" of the game or another player’s orders'' \cite[][p.~5]{C103}. However, later statements in the same paper specifically associated task \textit{execution} with agency, noting that the players who gave orders ``would like to have more \textit{agency} in the action (e.g. in shooting monsters, in rescuing people)'' \cite[][p.~8, our emphasis]{C103}. A paper grounded in Actor Network Theory first stated that agency arises from network effects, and is not the property of individual artifacts \cite[][p.~828]{C138}, but later discussed ``the agency of the artifacts'' \cite[p.~829]{C138}. 
Finally, some papers \cite[e.g.~,][]{C131, C122, C7, C96, C129, C131} explicitly indicated in the abstract or introduction that agency or autonomy was a significant themes of the paper but then never used the terms directly again. 

Very few papers explicitly engaged with existing theoretical accounts of agency and autonomy. However, in some papers that did, a certain understanding of agency or autonomy was implicit via the concepts and theories cited. Self-Determination Theory (including METUX), for example, was most frequently mentioned (n=17), particularly in works focusing on the player experience \cite[e.g.,][]{C38,C79,C141} or user experience \cite[e.g.,][]{C119,C134}. Research focusing on the immediate sense of control over actions often resorted to concepts from cognitive science, such as the notion of intentional binding \cite[e.g.,][]{C48,C62,C112}. Other conceptual frameworks mentioned include Actor Network Theory \cite[e.g.,][]{C127,C77}, value-sensitive design  \cite[e.g.,][]{C135}, social anarchism \cite{C104}, and Kantian philosophy \cite{C70}.

\subsection{Aspects of Autonomy and Agency}

A few papers in our corpus explicitly noted that agency and autonomy had ``many different facets'' \cite[][p.~12]{C70}. 
In line with this
we identified a variety of facets of autonomy and agency in the reviewed works. By analysing the corpus as a whole we found that these facets, and author' approaches to them, could be usefully organised around four broad aspects
: (1) A focus on \textit{self-causality} or \textit{identity}. (2) A focus on the \textit{experience}, or \textit{material} aspects of agency and autonomy. (3) The \textit{time-scale} over which autonomy and agency play out. (4) The degree to which
either \textit{independence} or \textit{interdependence} are emphasised. These aspects are summarised and defined in \autoref{tbl_rubric} and discussed in detail in the subsections below.

\subsubsection{Self-causality and Identity}

Accounts of autonomy and agency in our corpus contained elements of both \textit{causality} (related to the level and directness of the user's causal involvement), and  \textit{identity}: (related to the user's self and values). 
Nearly all discussions of autonomy and agency (n=159) dealt with causal issues to some degree, while just under one third (n=54) also dealt with issues of identity. We discuss these separately below.


\paragraph{Causality} \label{causality}
Causal aspects of agency and autonomy divided into cases of \textit{execution}: concerning active and direct material involvement in tasks and outcomes, and \textit{decision}: concerning choice and decision making. 
Discussions in terms of \textit{execution} varied considerably in complexity, and were found in the majority of papers (n=122).
Some papers focused on very limited 
interactions, for example,
the users' sense of  having caused 
individual events by simple interaction gestures such as pushing a button, with no element of personal choice or decision \cite[e.g.][]{C125, C128, C160, C110}.
Most papers focusing on execution approached agency and autonomy in a less atomistic manner:
focusing, for example, on
narrative choice \cite[e.g.,][]{C22} and control of action in video games \cite[e.g.,][]{C9, C38, C79, C80}
, people's involvement in everyday personal activities \cite{C43, C65}, or data analysis \cite{C89}.

Roughly two thirds of papers (n=99), 
discussed not only execution but also \textit{decision} and choice.
Often the nature of the activity under study made it hard to separate decision and execution (for example, agency in movement, directing attention, and sense-making in a museum \cite{C4}), but there were notable exceptions to this. Lazar et al., for example, distinguished ``the capacity to act'' from ``role in decisions'' \cite[][p.~1]{C19}. Similarly, Inman and Ribes linked these aspects of agency with different approaches to design, suggesting that "seamless" designs ``[grant] agency to users by lowering technical barriers of entry, [and] facilitating quick access to common operations'' \cite[][p.~2]{C71}, while "seamful" designs ``allow users to make up their own minds'' \cite[][p.~3]{C71}.
Meanwhile 36 papers focused significantly on decision or choice, giving little or no attention to execution.
In some of these cases the tasks under study (such as giving consent \cite{C44}) were decisional in nature, but
others discussed tasks with more substantial executional components, including gaming \cite{C141, C147}, and use of assistive technologies \cite{C70, C137}. 

Finally, some works clearly distinguished between
executional or decisional factors 
and saw each as having a different impact.
Two papers, for example, dealt with multiplayer games where decisional and executional roles were split between players, with one directing and the other acting. In both cases there was some ambiguity in how these aspects differently supported or undermined agency and autonomy \cite{C80, C103}. Karaosmanoglu et al., for example, stated both that ``agency/control [was] transferred to the'' player in charge of decision making \cite[][p.~10]{C80} and that in the executional role ``agency increased as they had a way to noticeably affect the game world'' \cite[][p.~12]{C80}. 
A number of other papers suggested that having a role in decision without involvement in execution resulted in diminished agency: Performing activities for oneself was seen to support agency and sense-of-self for people in supported living \cite{C19, C24}. Loss of executional agency was associated with loss of flexibility and adaptation both during supported communication \cite{C65}, and in interaction with intelligent technologies \cite{C3, C68, C73}.


\paragraph{Identity}
Some papers addressed
agency and autonomy primarily in terms of identity.
Again, these divided into two further categories: \textit{self-congruence} (n=41) which concerns the alignment of outcomes with users' values and goals, regardless of their involvement in decision or execution; and \textit{self-construction} (n=24) concerning effect on user’s identity, values, and goals.
Deterding \cite{C141}, for example, distinguished self-congruence from both decision and execution, stating that Self-Determination Theory ``does not equate autonomy with [...] the presence of choice [...] Choice is `merely' a facilitating condition that makes it more likely that individuals find a self-congruent activity'' \cite[][p.~3932]{C141}. 
However, despite this theoretical distinction, all but three \cite{C64,C77, C53} of the papers which emphasised autonomy 
in these terms
also retained some focus on executional and decisional aspects. 
One paper explicitly distinguished between aspects of causality and self-congruence of outcomes, 
and emphasised that the former supported the latter: ``as individuals express \textit{technical agency} by participating they can then advance their objectives in conversation, [...] \textit{colloquial agency}'' \cite[][p.~2, our emphasis]{C65}.
Often self-congruence of outcomes was emphasised in cases of active self expression by the user \cite[e.g.,][]{C3,C13,C145,C157}, or action and decision in pursuit of values and goals \cite[e.g.][]{C19,C22, C119}. 

Some papers suggested potential difficulties in ascertaining when activities and outcomes were self-congruent, since multiple conflicting goals and values might weigh on the same activity for a single user. 
Three papers referred to cases like this as ``paradoxes'' of agency \cite{C27,C29, C154}. In a study of how contexts support autonomy in gaming \cite{C141} one participant reported playing multiplayer game sessions a regular competitive gaming group (or ``clan'') during which he could not ``decide voluntarily to leave'' \cite[][p.~3935]{C141}. The author related this to autonomy by emphasising that this outcome was not congruent with their ``spontaneous desire'' \cite[][p.~3935]{C141}. However, given the player's wider commitment to social and competitive gaming, it seemed likely that the alternative outcome would not be congruent with \textit{longer term} values and goals.
Here both choices might be self-congruent, and different time-scales seemed important in distinguishing the impact of the different motives at play (see section \ref{time-scales} below for more discussion of these issues).

In 24 papers discussion of agency and autonomy concerned the genesis of users' values and goals --- what we termed issues of \textit{self-construction}.
As one paper put it: Autonomy is the ability ``to be one's own person, to be directed by considerations, desires, conditions, and characteristics that are not simply imposed externally upon one'' \cite[][p.~9]{C32}. Such papers focused on how technologies can ``shape the identity'' \cite[][p.~3]{C30} of users, whether or not this is intentional, or desired. 
Mentis et al.~suggested that assistive technologies which neglect psychological, cultural, and emotional aspects, risk ``reifying [disabled peoples'] dependency and their loss of self'' \cite[][p.~3]{C30}.
Other papers discussed more positive examples of self-deﬁnition and self-change:
Discussing, for example, the autonomous integration of social values in the context of self-control apps for children's video viewing \cite{C114},
 or how self-tracking systems for chronic health conditions could support agency by supporting users' reflection, self-appraisal, and self-awareness \cite{C126,C154}.


\subsubsection{Material and Experiential} \label{experiencematerial}
Papers in our corpus also addressed \textit{material} and \textit{experiential} aspects of agency and autonomy. We use ``material'' here to refer to both the material expression of autonomy and agency (e.g., as Coyle et al.~note: `` the fact of controlling an action'' as distinct from ``the immediate sense or experience of having done so'' \cite[][p.~2026]{C18}.), and also wider material factors which may impinge on this (e.g., being subject to coercion, lacking economic means, power or influence).
Papers across our corpus discussed both material and experiential aspects,
 though few explicitly distinguished between them.
Many papers (n=80) focused exclusively on material aspects of agency and autonomy. Such papers
discussed, for example, the material ability of people to act independently or under their own volition (e.g., support for personal mobility \cite{C43, C60}, communication \cite{C19, C25, C67}, or everyday living \cite{C21, C24, C88}), or the material ability to pursue one's own preferences and choices (e.g., at work \cite{C2, C50, C51, C90, C101, C108}, in social engagements \cite{C44}, or with respect to health \cite{C54, C58, C100}). 
A smaller number of papers (n=46) focused exclusively on experiential aspects --- for example, the sense-of-agency when triggering an event \cite{C158}, or the experience of autonomy while playing computer games \cite{C9}. 
Some papers discussed examples of material agency or autonomy with no clear experiential component: These papers focused on the autonomy and agency of organisations rather than individuals \cite[e.g.,][]{C83, C75}, and others drew on Actor Network Theory's account of agency --- a network effect of relations between human and non-human actors, sharply distinguished from intentionality \cite{C69, C77, C127, C138, C95}.

Finally, 33 papers discussed both material and experiential aspects. Here, it was sometimes emphasised that these aspects might not align with one another.
A number of papers indicated that the sense-of-agency in quite minimal interactions could be manipulated, by manipulating sense-of-ownership \cite{C128}, haptic feedback \cite{C116}, or the timing of events \cite{C125, C128}. Some noted that users can exhibit sense-of-agency with respect to situations where they have no causal influence \cite{C62,C125, C128}: 
Two papers showed that when the electrical stimulation of muscles was used to initiate the user's response to an event,
the user's experience of agency was increased by delaying the stimulation slightly (while still pre-empting the user's own response) \cite{C125, C128}
Several other papers drew on Self-Determination Theory, which (as discussed above) emphasises that sense of autonomy is supported by outcomes congruent with values \cite{ryan_autonomy_2004,ryan_brick_2019}, raising the \textit{possibility} that it may be manipulated without material change in the user's independence, range of choice, or influence on events \cite{C141}. However as noted above, \textit{in practice} 
in all these cases, material and experiential aspects of agency or autonomy did not diverge.

\subsubsection{Time-scales} \label{time-scales}
Aspects of autonomy and agency were often differentiated by the time-scales of activities and experiences. 
We found that papers addressed three broad time-scales: \textit{micro-interactions} (autonomy and agency experienced or exercised over a few seconds or less), \textit{episodes} of interaction (seconds to hours), or autonomy and agency at longer time-scales in \textit{life} (days to years). The character of autonomy and agency, and authors' approaches to them differed significantly between these three time-scales. While some features --- such as a focus on self-causation --- were consistent across all scales, other features differed significantly.
While these differences were apparent over the corpus as a whole, only a few papers \textit{explicitly} addressed time-scales distinctions, or addressed issues of autonomy and agency at more than one time-scale. Some however, did address issues of time-scale, noting for example that ``immediate autonomy ... can have an effect on events in in the distant future; and vice versa, long-term decisions can have an impact on the very present.'' \cite[][p.~3]{C70}. Such works pointed to a range of tensions, trade-offs, and synergies across time-scales. 

72 papers focused purely on \textit{episodes} of interaction --- shorter than a day in length. 
All of these 
addressed issues of 
self-causality (i.e., execution of actions (n=60) and decision or choice (n=42)). Relatively few (n=13) also discussed identity-related aspects of agency and autonomy. Substantial discussion of 
self-congruence and identity
was mostly limited to a few studies 
in which users were interviewed about
their daily lives, or which made use of observations and experiments in the field \cite[e.g.,][]{C71, C26, C20, C114}. While these papers still focused on episode-level aspects of autonomy and agency, the studied episodes were embedded in participants' everyday lives, allowing results to be informed by this. For example, Deterding discussed how individual episodes of life-situated video-game play were impacted by contextual factors, such as the need to ``make time'' for play by clearing other tasks \cite{C141}. Other examples deployed apps for emotional communication to families and couples \cite{C3} for a number of weeks, or combined lab-based experiments on the use of assistive communication-devices with user interviews \cite{C65}.

Meanwhile, 35 papers focused solely on time-scales in wider \textit{life}.
Again, all these papers focused to some degree on issues of self-causation, whether executional (n=25) (e.g., ``re-interpreting [a generated conversational prompt] to take the lead in a conversation'' \cite[][p.~8]{C8}) or decisional (n=20) (e.g., decisions about parenting in divorced families \cite{C75}). Half of these papers (17) also discussed identity-related aspects of autonomy, such as
how assistive technologies ``shape the identity of the very people they are to help'' \cite[][p.~3]{C30}. Some of these papers indicated that length of engagement might have implications for agency and autonomy; via the potential to imbue technologies and places with meaning over time \cite{C77}, or via habit formation and reflective self-change \cite{C45}.  

Finally, a number of papers addressed ``micro-interactions'', under a few seconds in length (n=13) dealing with the \textit{experience} of self-causality while triggering individual events 
\cite{C125, C128, C160, C110}. This work focused exclusively on this very short time-scale, isolating agency in \textit{execution} from issues of \textit{decision}, \textit{self-congruence}, and more generally from any wider interaction context. 
Six of these papers \cite{C12,C18,C47, C62,C112, C159} focused on the specific neuro-scientific construct \textit{sense-of-agency}, which refers to the \textit{implicit} sense of authorship of an action. The operationalision of this via millisecond-level variations in a reaction-time metric --- so-called \textit{intentional binding} --- was seen as promising for understanding ``the awareness
of owning the actions’ outcomes'' \cite[][p.~2426]{C47}, and ``how people
experience interactions with technology'' \cite[][p.~2025]{C18}. However, two recent papers in this group noted the uncertain relationship of this measure to other constructs of agency, and their results indicated that the relationship of temporal binding to conscious user experience remained unclear \cite{C12, C159}. 

\paragraph{Relationships between time-scales}
Just under a quarter of papers (n=41) discussed autonomy and agency in both short \textit{episodes} of interaction, and also on longer time-scales.
In most examples, one time-scale or the other received only cursory discussion. Wan et al.~\cite{C24}, for instance, focused on day-to-day agency of people with dementia, though in the course of this they described some shorter episodes in which agency was at issue. However, some works
addressed
tensions and trade-offs between agency and autonomy at different time-scales. 
Three papers described apparent ``paradoxes'' \cite{C27,C29, C154} whereby restricting agency or autonomy also seemed to increase agency or autonomy. 
These cases spanned a range of contexts --- from digital self-control \cite{C27} to spousal surveillance in dementia care \cite{C29}, and in all cases, the restricted and supported aspects of agency were characterised by different time-scales: participants accepted restrictions in \textit{episodes} of interaction in order to gain agency or autonomy in their wider life. For example, users blocked immediate access to digital distractions to achieve longer-term personal goals \cite{C27}, and people with mild cognitive impairment accepted blocks or vetos on access to risky websites to retain their longer-term freedom to browse the web independently \cite{C29}.

As well as trade-offs and tensions, some papers indicated synergies 
across different time-scales. 
Several papers described how episodes of executional agency in crafting, caring for others, or simply performing everyday activities independently, could support agency in self-definition in wider life. For example, in the context of digital social sharing for people with dementia, Lazar et. al referenced arguments that ``low-level, proprioceptive control over one’s environment  is part of the creation of agency'' \cite[][p.~2157]{C19} in wider life.

\subsubsection{Independence or Interdependence?}
Lastly, several papers placed individual independence at the centre of their discussions of autonomy and agency (n=50). In line with a statement by Güldenpfennig et al., we found that  ``autonomy'' was often ``translated to ``independence'' without any further differentiations'' \cite[][p.~13]{C70}. Meanwhile, we found that other papers emphasised \textit{inter}dependence (n=54): focusing on how social contexts support agency and autonomy, and provide the necessary horizon against which these ideas must be understood. 
This tension between independence and interdependence was most notable in discussions of ``autonomy'', though also present in some discussions of ``agency''.
In some work ``autonomy'' appeared as a near synonym for ``independence''. For example, Garg noted that ``autonomy-related changes occur in the relationship between parents and children ..., as teenagers try to assert their independence'' \cite[][p.~1]{C161}, and Partala defined autonomy as ``to actively participate in determining own behavior without external inﬂuence'' \cite[][p.~788]{C134}.
Along similar lines a number of papers addressed \textit{trade-offs} between independence and interdependence: emphasising how the role of others could undermine independence and thereby agency or autonomy. In a cooperative gaming scenario, for example, it was suggested that ``tight dependence on each other led some players ... to report a lack of autonomy'' \cite[][p.~11]{C103} (note though that no comparison was made to a version of the game in which these roles were more independent).

Other works discussed independence with regards to barriers to agency and autonomy which followed from health conditions, or other material factors.
Here, independence was emphasised insofar as there was a risk that it could be lost, as subjects were impeded in some way from acting independently for themselves, resulting in meaningful impacts on their lives. One paper, for example described work ``to make board games accessible, ensuring the autonomy of players with visual impairment'' \cite[][p.~2]{C14}, and other papers discussed autonomy of mobility \cite{C105, C24}.
However, such focus on independence was often situated against a wider backdrop of interdependence \cite{C70, C141, C46, C149}. Characteristic of this, one paper noted that while people might wish to ``accomplish as much independence as possible ... social networks contribute significantly to every person’s autonomy and welfare'' \cite[][p.~10]{C70}, and that ``the liberal-individualist account of autonomy over-emphasizes physical independence and does not sufficiently recognize the inter-dependency of all people'' \cite[][p.~128]{C70}.
Elsewhere, papers described ways in which interdependence might support agency and autonomy: one paper found that individuals' sense of agency in online posting was boosted when they saw evidence of others' online activity \cite{C13}, and several papers noted the crucial role played by contexts-of-living in determining the autonomy outcomes produced by deployed technologies \cite{C46, C24, C58, C127, C137, C145}.

\section{Discussion}

Our results demonstrate a consensus among HCI researchers that there is value in supporting human autonomy and agency, and that these concepts are key to fundamental questions about user experience and the impact of technologies on identity and personhood.  
However, behind this consensus we find considerable diversity in \textit{how} these concepts are valued, and even in understandings of \textit{what} agency and autonomy entail. 
As such, our findings indicate that autonomy and agency currently function as resonant, albeit vague, \textit{umbrella concepts}: gathering together a wide range of perspectives on what may or may not be related phenomena \cite{hornbaek_commentary:_2018}. 
Indeed, our analysis revealed structure and meaningful distinctions running across the corpus: specifically, concerning the time-scales of behaviour; the focus on experiential or material issues; the focus on causal or decisional involvement, or on issues of identity; and in terms of how strongly independence was emphasised. 
However, we found these distinctions mostly operated implicitly. It was rare for authors to explicitly articulate particular aspects of agency and autonomy, let alone discuss coordination, tensions, and trade-offs between them.

Previous work has argued that such vaguely defined umbrella concepts ``challenge our ability to accumulate and communicate knowledge'' \cite[p.~2]{tractinsky_usability_2018},
and we found evidence of this in our corpus.
Within particular domains, we found some evidence of authors building upon one another's findings (e.g., dementia care \cite[e.g.,][]{C70}, low-level agency in minimal interactions \cite[e.g.,][]{C12}). However, we found few cases where work built on findings outside its own immediate domain or context. This was the case despite evidence of considerable commonalities between research in different contexts, -- and points to missed opportunities. For example, we found papers dealing with digital fabrication \cite{C68} internet-of-things \cite{C50}, and accessible technologies \cite{C70} all addressing how to balance executional and decisional aspects of autonomy and agency when delegating to technology and other people. It seems likely that there are opportunities for each of these communities to learn from the findings of the others.

One response to this situation might be 
to abandon efforts to maintain and develop these wider understandings of agency and autonomy
and instead isolate particular aspects as a series of distinct, well-defined, and independent constructs
\cite[e.g.,][]{tractinsky_usability_2018}. However, in line with previous discussions of the `usability' construct \cite{hornbaek_commentary:_2018}, we do not see this as the best way forward. We suggest that the problem is not simply that autonomy and agency are umbrella concepts. Many important scientific constructs --- including  affect, emotion and socio-economic status --- function in this way, gathering together a complex of different perspectives and subconcepts. While this brings limitations in treating the concepts as unified unitary concepts, this can still support communication and the identification of important research topics \cite{hornbaek_commentary:_2018}. Instead, we argue that,
in addition to an occasional local lack of specificity and precision,
a larger problem is that HCI currently \emph{lacks coordination between different approaches to agency and autonomy}. 
We suggest there is value in developing existing approaches to autonomy and agency: partly by clarifying individual aspects and local understandings, but also by clarifying the relationships between these aspects, and coordinating understandings and issues across contexts and communities in HCI; to help autonomy and agency function as \textit{boundary objects}.

\subsection{Clarifying Agency and Autonomy as Boundary Objects} 
Boundary objects are flexible yet 
robust concepts that can help coordinate the perspectives and activities of different communities of practice \cite{nuchter_concept_2021}, 
without requiring strict consensus on precise definitions \cite{leigh_star_this_2010}.
Previous work has emphasised that, to function well, boundary objects should be sufficiently flexible to meet the informational and communicative needs of the communities involved \cite{leigh_star_this_2010}. Our review suggests 
that agency and autonomy already fulfil this criterion, as illustrated by their continuing significance in a range of different issues and contexts, over three decades of HCI research.
However, while such interpretive flexibility is a well known property of boundary objects \cite{velt_translations_2020}, Star and others have emphasised that interpretive flexibility is not sufficient on its own \cite{leigh_star_this_2010}; boundary objects can fail to function well, 
if they are not also robust \footnote{Brian Cantwell-Smith suggests that this robustness might consist in some degree of consistency in the ontologies of the multiple communities involved, but points to the difficulty in specifying this further without falling into naïve realism \cite[pp.~219-227][]{bowker_boundary_2015}. A similar problem is addressed in Hasok Chang’s recent pluralist scientific realism by focusing on what he calls the "operational closure" of theories and the manner in which they are \textit{constrained} by reality \cite[p.~169][]{chang_realism_2018}. While we believe these resources can inform future discussion of coordination between communities in HCI, that discussion is beyond the scope of this paper.} enough in usage to retain a broader identity across interpretations. This is required to support coherence across intersecting communities \cite{star_institutional_1989}. 
For example, in sustainability studies, it was found that the concept of ``resilience'' did not succeed as a coordinating boundary object  \cite{nuchter_concept_2021}, since different communities' understandings did not overlap in important ways. 

What do robustness and flexibility mean in practice? Star \cite{leigh_star_this_2010} emphasises that  boundary objects rely on both 1) the refinement of the concepts to address local concerns, and 2) the coordination of these local understandings with the concept's wider identity. In terms of autonomy and agency in HCI, we suggest this might mean 1) \textit{local} attention by researchers on the aspects of autonomy and agency which are relevant to concerns in their area of work, and 2) \textit{integrative} work to relate these aspects to one another, and to situate local findings and understandings in the context of wider understandings. 
Our analysis points to concrete ways in which the HCI community might approach these two activities, to develop autonomy and agency into better functioning boundary objects:

\subsubsection{Articulate 
Locally Relevant Aspects of Agency and Autonomy}
 First, to advance \textit{focused} local work within particular communities and contexts, there is a need to be more explicit in identifying, defining, 
and clarifying 
the aspects of agency and autonomy which are at issue in particular areas of HCI research. Our findings indicate that currently only a minority of papers provide explicit working definitions of agency and autonomy that are specific to the work carried out. Fewer still articulate distinctions within these definitions that are relevant to their work. In future, more work might follow the example of papers in our corpus \cite[e.g.,][]{C65,C70}
and clarify issues in their domain by articulating distinctions between aspects of agency and autonomy: for example, between decisional and executional autonomy \cite{C70}, or between technical 
(closely related to our our category of \textit{execution}) and colloquial agency (mapping to our categories of decision/self-congruence) \cite{C65}. 
Such distinctions were operative in much work in our corpus, but left implicit. Articulating them explicitly could provide a common vocabulary to help coordinate efforts, identify design strategies, sensitise researchers to relevant issues, and support researchers in identifying relevant work. 

The aspects we articulate in our review -- self-causality and identity; experiential and material; time-scales; independence and interdependence, -- offer one potential source of such distinctions, grounded in existing work in HCI, and sometimes (as in the case of decisional and executional aspects \cite[e.g.][p.~3]{C70}) directly echoing categories which are already in use. Note that these aspects are considered neither definitive nor exhaustive. Nonetheless, we found that they captured important distinctions across 30 years of HCI research. 
We further suggest that they are sufficiently flexible and descriptive 
to avoid 
bias towards any particular theoretical perspective or approach. 
The METUX framework offers another category of distinction, describing seven \textit{spheres} of life which are expected to be relevant to issues of wellbeing \cite{peters_designing_2018}, and though we suggest 
this lacks the granularity of our four aspects, there may be cases in which a single-dimension is adequate and even preferable.
More broadly, Self-Determination Theory --- the theoretical basis of METUX --- offers a wide range of theoretical resources for understanding autonomy on different scales. 
At present only limited aspects of SDT have found application in HCI (largely SDT's Basic Needs Theory component \cite{ryan2017basic}, and certain standardised questionnaires). Future research might focus on underused resources such as Organismic Integration Theory \cite{ryan2017organismic}, and Cognitive Evaluation Theory \cite{ryan2017cet} which might bring clarity to understandings of longer term autonomy, users' integration of values, and wider contextual factors which might impact on autonomy \cite{ballou2022self,tyack_self-determination_2020}.

\subsubsection{Investigate Relationships Between Different Aspects}

Future research might also give more attention to \textit{integrative} work: understanding how particular aspects of autonomy and agency relate to one another, and how local understandings fit into the wider account of agency and autonomy in HCI. While focused work on isolated aspects of agency and autonomy is currently common (if often lacking in explicit definition), very little work attempts to integrate and relate different understandings and aspects of agency and autonomy.
First, we found that it is rare for papers to address issues of autonomy and agency on more than one broad time-scale. 
This leaves a range of open questions: Does support for agency and autonomy in short episodes of interaction impact upon wider life (e.g., the potential ``contradictory parallel effects'' highlighted by the METUX framework \cite[p.~7]{peters_designing_2018})? 
Conversely, do episodes of autonomy and agency in the lab function the same way as similar episodes when they are situated in participants' everyday lives? 
Does that wider context of autonomy experience in life override, or otherwise modulate, episode level experience? 
Some papers in our corpus provide a model for addressing such questions by studying episodes of agency situated in participants' wider lives \cite[e.g.,][]{C71, C26, C20, C114}, or using field studies and interview studies to contextualise results in lab results. Another avenue is to leverage the wider resources of Self-Determination Theory (discussed above) to understand how other longer-term factors --- such as value integration and contextual factors --- impinge on episode-level experiences of autonomy. The recently validated User Motivation Inventory might be useful here \cite{bruhlmann_measuring_2018}; as might be the example of Tyack and Wyeth's work on autonomy in gaming \cite{tyack_small_2021}, which
augments single-session questionnaire-based studies with semi-structured interviews to address autonomy experience in everyday contexts of play. 

Beyond time-scale, our findings show that other aspects of autonomy and agency 
were also mostly dealt with in isolation. Several papers focused on low-level sense-of-agency in micro-interactions \cite[e.g.][]{C125, C128, C160, C110}, operationalised via a metric called \textit{temporal binding} 
(i.e., a reaction-time based measure, 
which is expected to correlate with the user's \textit{implicit} sense of having caused an action \cite{moore_intentional_2012}). 
To date HCI research has focused on this construct in isolation from factors such as decision, self-congruence of outcomes, and social context. This is an understandable limitation in early work. However, our review shows the important role of these factors in HCI's understandings of agency and autonomy, and recent work in cognitive science indicates that these factors (in particular social \cite{vogel_temporal_2021, malik_social_2019} and personality factors \cite{hascalovitz_personality_2015}) may affect temporal binding. One route to address this in future work is to investigate the relationship between temporal binding and predictions drawn from Self-Determination Theory's account of autonomy \cite{ryan_autonomy_2004}.
This might, for example, mean introducing meaningful (self-congruent or incongruent) choices into existing experimental paradigms. Work in this direction might also help clarify open questions around how temporal binding measures relate to UX \cite{C159}.

Finally, we suggest there is value in continuing wider inter-community efforts to coordinate understandings --- extending the previous CHI workshops on autonomy \cite{friedman_user_1996,calvo_autonomy_2014}, and our own work in this paper. Such work can serve to take stock of the interrelations between different communities within HCI for whom autonomy and agency are important concepts. If such events and projects are framed in a sufficiently pluralistic manner, they can support the ``tacking back-and-forth'' \cite[][p.~605]{leigh_star_this_2010} between local and wider community perspectives necessary to support the maintenance of boundary objects: helping articulate potentially valuable consistencies and commonalities in approach.

\subsection{Clarifying how Aspects of Agency and Autonomy Relate to Outcomes} 

While many of the reviewed works associated 
autonomy and agency with a wide range of positive outcomes, we often found a lack of specificity in \emph{how} these outcomes might be linked to particular expressions of autonomy and agency.
This was least problematic in work which pointed to relatively modest conclusions, e.g., linking outcomes readily measurable at episode level \cite[e.g., task performance,][]{C99} to autonomy and agency during single-episodes of interaction. Here, even if it was not directly stated which aspects of autonomy and agency were considered relevant, this was often evident via the nature of experimental manipulations \cite[e.g.,][]{C5}, or via questionnaires which addressed well-defined operationalisations of autonomy \cite[e.g.,][]{C113}. 
However, some such scenarios raised important questions.
For example, in work which operationalised sense-of-agency via temporal binding \cite[e.g.,][]{C159}, the association between temporal binding and UX was generally asserted rather than directly tested. Only in more recent work was the relationship between temporal binding and user experience directly examined, and ultimately found to be somewhat unclear \cite[e.g.,][]{C159}.

Elsewhere the relationship between agency, autonomy and UX was unclear in other ways. In some instances, autonomy was treated as an \textit{antecedent} of good UX \cite[e.g.,][]{C5, C18}; in others as a \textit{component} of good experience, and a positive experiential phenomenon in its own right \cite[e.g.,][]{C38}.
This suggests that the UX community lacks agreement on clear models of how these constructs relate to one another. In line with this, the links between autonomy or agency and UX were often articulated in quite general terms. Very few papers investigated how autonomy and agency related to specific components of user experience, such as sense-of-presence \cite{C85}. Future work might aim to move beyond the current understanding of agency and autonomy as components of generically positive experience, and instead seek to better understand how and why specific aspects of autonomy and agency foster specific aspects of positive experience.

Outside UX and player experience, papers often connected support for agency and autonomy to ``high value'' outcomes, such as sense-of-self and well-being \cite[e.g.,][]{C27}. However, this connection was always drawn on the basis of prior work in other domains, and not explicitly examined in the case of technology use. Again, reviewed works often left it unclear which aspects of autonomy and agency were expected to lead to these positive outcomes. Exceptions to this were found in some qualitative field work, addressing for example, how low level ``proprioceptive'' agency in execution could be expected to lead to higher level agency outcomes and well-being \cite[p.~2157]{C19}. Or in work that reported on the complex trade-offs in well-being support between what we have characterised as execution, decision, and self-congruence \cite{C70}.
However, as we have noted, autonomy and agency were generally studied in single sessions of interaction, without addressing the user's wider life context, where high-value outcomes like well-being, motivation, and sense-of-self play out. In fact, our corpus contained no works that compared such outcomes pre- and post-session, or observed it over long term studies. Of course, it is challenging to address long-term outcomes in HCI studies and interventions \cite{kjaerup2021longitudinal}. However, without such work, it is hard to see how we can meaningfully investigate the impact of agency and autonomy on outcomes such as well-being and sense of self. In addition to the qualitative work discussed above, we suggest that the wider resources of Self-Determination Theory, beyond Basic Needs Satisfaction might be drawn upon here, and that work by Tyack and Wyeth provides an exemplary and achievable model for such future work \cite{tyack_small_2021}.

Another complex question raised by our analysis concerns how we can disentangle experiential and material aspects of autonomy and agency and their impact on outcomes. As reported in section \ref{experiencematerial}, we found suggestions that some experiential and material aspects could be decoupled. For example, that the experience of autonomy and agency can be altered without affecting the user's causal involvement or material control over outcomes \cite[e.g.,][]{C141}. Similarly, it was found that sense-of-agency over sensorimotor actions could predict training outcomes even where users had no causal role in the outcome \cite{C125}. 
These examples indicate that there are clear opportunities for exploitation of such gaps -- sometimes to the detriment of the user: for instance, using the feeling of autonomy or agency as an alibi for materially reduced efficacy, choice, or scope of control. This is illustrated in our corpus in work by Bakewell et al., where companies granted workers self-management in certain aspects of their working life in a way which created wider, and sometimes undignified, restrictions on behaviour and efficacy \cite{C90}. 
Based on our findings, one straightforward recommendation for future work is to be (more) attentive to this distinction between experiential and material aspects of autonomy and agency. It is crucial that HCI researchers clearly report the relevant aspect(s) of autonomy and agency, and focus analysis appropriately, particularly where technology is likely to impact outcomes meaningful to the user. Future work might also seek to understand whether --- as predicted by SDT \cite{ryan_autonomy_2004,ryan_brick_2019} --- the congruence of outcomes will impact on autonomy independently of changes in, e.g., available choice, executional involvement, or the user's independence of action. For example, the degree of control afforded in game and VR environments may make them a suitable vessel for such work, though this will require careful study design.






\subsection{Ethical Challenges} 
Finally, our analysis raises ethical challenges.
First, there is the question of who benefits from the user's agency or autonomy. Our corpus contained examples where these benefits primarily accrued to others --- employing organisations, or sellers of products and services. 
This might not always be a problem in itself. One paper, for example, reported that delegation of worker decision-making to AI reduced attentiveness and engagement, leading to poorer task outcomes \cite{C94}. In this case while the organisation is the primary beneficiary of the worker's agency, it is not immediately clear that this is problematic for the worker.
However, other papers gave examples where benefits and burdens were more problematically distributed: users gained only limited and localised control or choice, or only the experience of agency and autonomy, while in return accepting increased burdens and material restrictions on aspects of behaviour \cite[e.g.,][]{C90}. 

One key problem here is that autonomy and agency are resonant words, associated with a wide range of meanings. 
Sometimes, autonomy and agency are ``human rights'' \cite[e.g.,][]{C65,C135} or are meaningfully associated with well-being \cite[e.g.,][]{C39,C102}. In other cases, agency and autonomy have more limited ramifications, making it more acceptable to treat them as manipulable parameters which can improve work outcomes \cite[e.g.,][]{C90, C94} or the appeal of our products \cite[e.g.,][]{C5, C20}. 
This ambiguity can allow vague and under-specified claims of autonomy support to serve as cover for exploitative or manipulative practices. 
Clarity in addressing agency and autonomy therefore has not only practical implications for HCI research, but also ethical importance.
A simple step towards addressing this is to spell out (expected) outcomes; not every study addressing autonomy and agency can plausibly attach its outcomes to the most lofty ramifications and values of these concepts. 
For example, while agency has been hypothesised as the ground of identity and the self-other distinction \cite{KNOBLICH2003487}, the relevance of this to agency during button pushing seems limited. Instead, authors should focus on plausible, direct, outcomes which are relevant to the time-scale and scope of the work carried out, and relevant to the specific aspects of autonomy and agency involved. Publication pressures can easily drive us to inflate the value of our work, but in the case of autonomy and agency there seems a concrete risk in this value-inflation: it can provide a vague, peer-reviewed alibi for those who might exploit humans under the banner of apparent autonomy support.


Another ethical question raised by our analysis concerns 
autonomy in self-construction, and how we should handle the delicate issue of becoming involved in the user's processes of self-change, and self-control. This seems particularly relevant for behaviour change technologies and technologies which support reflection.
Across our corpus, three papers addressed different contexts of self-change: helping children internalise behavioural norms \cite{C114}, supporting trafficked women to understand and address their exploitation by others \cite{C109}, and helping abusive men take responsibility for their behaviour \cite{C154}. 
These particular examples stand out as being carefully conducted, and in each case working towards outcomes likely congruent with users' long-term goals and values. 
However, these cases also illustrate the delicate balance involved in such work. The third example \cite{C154}, in particular, involved a degree of short-term thwarting of autonomy, intentionally restricting meaningful choices to encourage the men to reappraise their identities and responsibility.
Such work requires careful reasoning about what aspects of autonomy and agency will be restricted and supported, on what time-scales, and with what consequences. It requires reasoning about whether the long-term outcomes can be expected to be congruent with the user's values and goals, and if not, then what warrants the intervention.
Not all cases of persuasive technology are so dramatic and significant as the above examples, but the issues raised here can be informative elsewhere.
One special case, for example, concerns apps for self-control, where user's goals, values and intentions may vary over time, and care will be required in deciding how, and how much, to hold users accountable to their past intentions \cite{LYNGS2022102869}.
Again, we suggest that one route to the required care in such cases is to understand the different scales and aspects of autonomy and agency involved, and clarity about what is supported and thwarted, when and where. 
Prior work that deals with such trade-offs, and with the warranting of agency delegation \cite[e.g.,][in our corpus]{C50, C70}, can be useful in guiding reasoning in such situations.
Thinyane and Bhat's work on the support of trafficked women \cite{C109} offers a sophisticated discussion of relevant issues, articulated with reference to Sen's Capabilities framework \cite{sen_inequality_1995}. Likewise, we see much promise in future work developing guidelines specific to user autonomy and agency in persuasive and reflective technologies, and more broadly in technologies for self-change.

\subsection{Limitations and Open Questions} 

Addressing concepts of such complexity inevitably means that certain issues must be left out of scope. First, this review focuses on agency and autonomy, and does not consider any of the agency-adjacent terms which have proliferated over the past few decades, such as ``empowerment'' \cite{schneider_empowerment_2018}, ``efficacy'' \cite{bandura_human_1991}, or ``competence'' \cite{bauer_competence_2000}. Agency and autonomy seemed to us to be the primary concepts here, whereas the other concepts can often be considered special cases, derivatives, or complexes which involve aspects of autonomy and agency alongside other factors.  Moreover, ``empowerment'' is already the subject of a recent review at CHI \cite{schneider_empowerment_2018}. That said, fitting these adjacent constructs into the overall landscape of agency and autonomy (e.g., by drawing on the four aspects outlined in this review) could provide useful insight.
Second, the scope of the review --- spanning over 30 years of research --- means that we have addressed agency and autonomy at a relatively high level. This is in line with the goals of this paper: to provide a view of how these concepts and their accompanying issues are understood and can be better coordinated across HCI. However, future reviews which address understandings of these themes in specific domains and areas of research would help clarify understandings in particular communities and contexts, as well as furthering the development of agency and autonomy as boundary objects for HCI.


\section{Conclusion}

This paper presents a review of 161 papers which address issues of agency and autonomy, over 32 years of HCI research. Our analysis 
identifies both the high value of these concepts for HCI research, their flexibility, and the degree of ambiguity in how they are understood. We find that, at present, these terms are treated as umbrella concepts -- broad concepts, subsuming a surprising diversity of understandings and theoretical underpinnings. 
The terms are largely used interchangeably and given a wide range of different meanings. This makes it difficult to address these concepts as a whole, leaves it unclear how different understandings of autonomy and agency relate to each other and to particular outcomes, and may impede researchers in identifying and building upon relevant prior work. 

To address this situation our analysis identified four aspects which help clarify understandings of agency and autonomy in HCI:
1. issues of \textit{self-causality} and personal \textit{identity}; 2. the \textit{experience} and \textit{material} expression of autonomy and agency; 3. particular \textit{time-scales}; and 4. emphasis on \textit{independence} and \textit{interdependence}.
These aspects may guide researchers to relevant prior work, and help situate their own work in the landscape of research on these concepts. 
We also point to future work which can develop agency and autonomy as boundary objects for HCI: constructs which coordinate the perspectives of various communities of practice \cite{leigh_star_this_2010}, driving design and understanding forward across multiple domains. To this end, we outlined avenues for HCI researchers to both clarify local understandings of autonomy and agency, relevant to particular issues, and coordinate these with wider community understandings. Specifically, we recommend that HCI researchers (1) explicitly state definitions and understandings of these concepts. (2) Leverage the four aspects articulated in our review to specify aspects of agency and autonomy that are relevant to their concerns. (3) Pursue greater specificity in linking particular aspects of autonomy and agency to outcomes. Finally, we call for (4) more integrative work, both to understand how different aspects of autonomy and agency interrelate, and to identify commonalities across different HCI communities and domains.

\begin{acks}
Funded by the European Union (ERC, THEORYCRAFT, 101043198). Views and opinions expressed are however those of the authors only and do not necessarily reflect those of the European Union or the European Research Council Executive Agency. Neither the European Union nor the granting authority can be held responsible for them.
\end{acks}

\bibliographystyle{ACM-Reference-Format}
\bibliography{references, additional_refs, corpus}

\end{document}